\documentclass[pss]{wiley2sp} 
\usepackage{amsmath}
\usepackage{bm}              
\usepackage{w-greek}         

\begin{document}

\title{Topological insulators from the Perspective of first-principles calculations}

\titlerunning{A review of topological insulators }

\author{%
  Haijun Zhang,
  Shou-Cheng Zhang\textsuperscript{\Ast}}

\authorrunning{Haijun Zhang et al.}

\mail{e-mail
  \textsf{sczhang@stanford.edu}, Phone:
  +01-650-723-2894, Fax: +01-650-723-9389}

\institute{%
  Department of Physics, McCullough Building, Stanford University, Stanford, California 94305-404531}
%

\received{XXXX, revised XXXX, accepted XXXX} 
\published{XXXX} 


\abstract{%
%
%
%
\abstcol{%
Topological insulators are new quantum states with helical gapless edge or surface states inside the bulk band gap.
These topological surface states are robust against the weak time-reversal invariant perturbations, such as lattice distortions and non-magnetic impurities. Recently a variety of topological insulators have been predicted by theories, and observed by experiments. First-principles calculations have been widely used to predict topological insulators with great success.

}
{%
In this review, we summarize the current progress in this field from the perspective of first-principles calculations. First of all, the basic concepts of topological insulators and the frequently-used techniques within first-principles calculations are briefly introduced. Secondly, we summarize general methodologies to search for new topological insulators. In the last part, based on the band inversion picture first introduced in the context of HgTe, we classify topological insulators into three types with \textbf{s}-\textbf{p}, \textbf{p}-\textbf{p} and \textbf{d}-\textbf{f}, and discuss some representative examples for each type.

}}

%
%
\titlefigure[height=7cm,width=6cm]{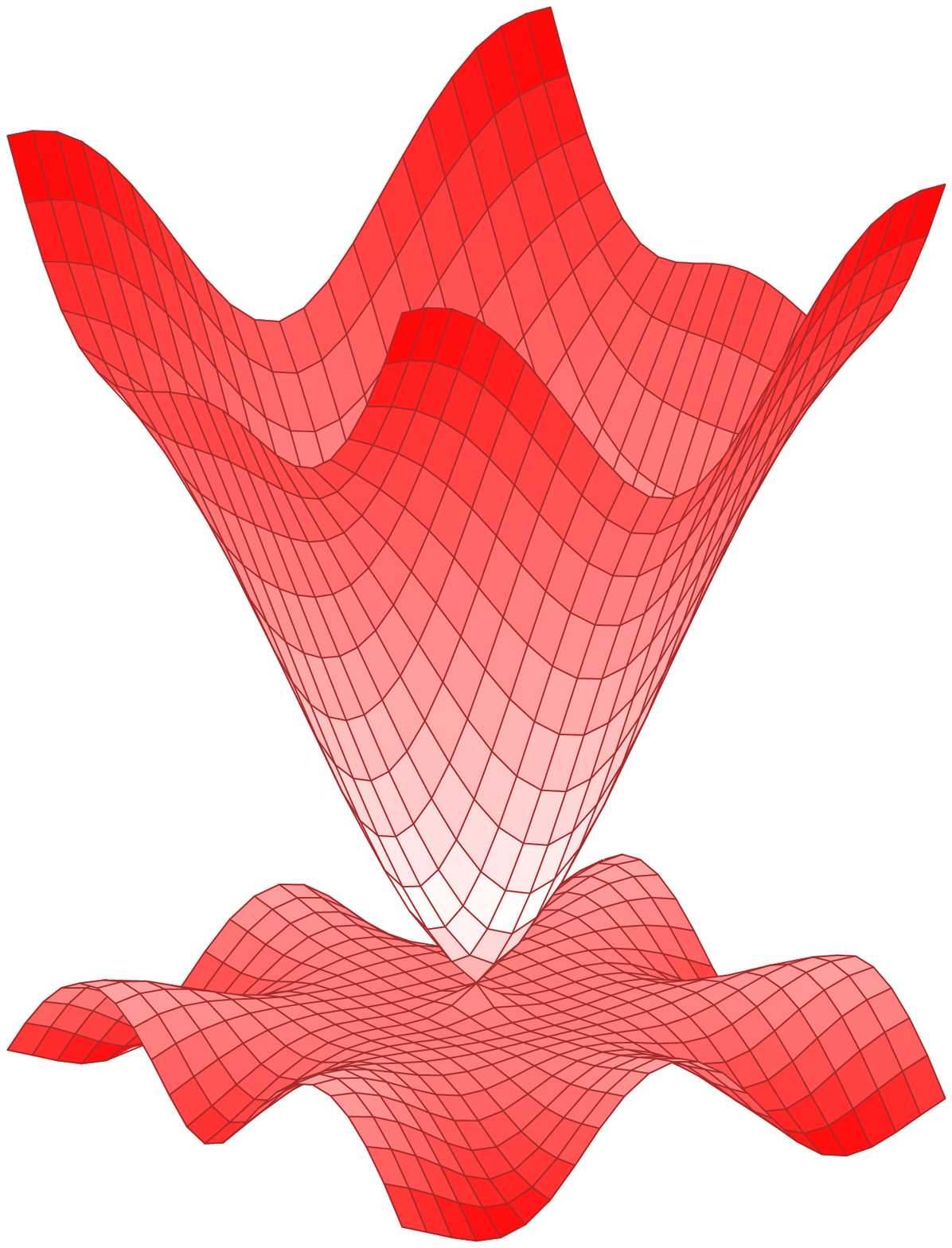}
\titlefigurecaption{%
  Surface states of topological insulator Bi$_2$Se$_3$ consists of a single Dirac cone, as obtained from first-principles calculations.}

\maketitle   

\section{Introduction}
In two-dimensional electron systems at low temperature and the strong magnetic field, the Hall conductance $\sigma_{xy}$ takes quantized values\cite{klitzing1980}, which proved to have a fundamental topological meaning\cite{thouless1982}. $\sigma_{xy}$ can be expressed as an integral of the first Chern number over the magnetic Brillouin zone. Recently quantum spin Hall(QSH) state was predicted and observed in CdTe/HgTe quantum well\cite{bernevig2006d,koenig2007}. In this system  time-reversal symmetry(TRS) is present, and spin-orbit coupling(SOC) effect plays the role of Lorentz force in QH effect. The concept of QSH can be generalized to three-dimensional(3D) topological insulators with TRS\cite{fu2007b,qi2008}. The electromagnetic response of a topological insulator is described by the topological $\theta$ term of $S_{\theta}=(\theta/2\pi)(\alpha/2\pi){\int} d^3xdt \textbf{E}\cdot\textbf{B}$ with $\theta=\pi$ where \textbf{E} and \textbf{B} are the external electromagnetic fields\cite{qi2008}. This indicates the physically measurable and topologically non-trivial response, which opens the door for experiments and potential applications of topological insulators.

Both 2D (QSH) and 3D topological insulators have interesting physical properties\cite{Qi-physics-today2010,Moore2010,Hasan2010,Qi2011}. In this review we focus on 3D topological insulators with TRS. In this field, an important task is to systematically search for all topological insulators. In this process, first-principles calculations played a crucial role. Up to now, most of topological insulators were predicted first by first-principles calculations, and observed subsequently by experiments.

\section{Theories and Methods}
\subsection{First-principles methods}
Density functional theory(DFT) is a formally exact theory based on the two Hohenberg-Kohn theorems(HK)\cite{hk1964}, but the functional of the exchange and correlation interaction is unknown in Kohn-Sham(KS) equation\cite{ks1965}. In order to do numerical calculations, the local-density approximation(LDA)\cite{ks1965} and Generalized Gradient Approximation(GGA)\cite{GGA1983,GGA1988} are usually used to approximate the exchange and correlation interaction in KS equation. Based on recent experiences, LDA and GGA work quite well for the study of topological insulators, because most topological insulators found to-date are weakly correlated electronic systems.

As we know, the conventional LDA and GGA first-principles calculations tend to underestimate the band gap\cite{Perdew1983,Sham1983}. However the band gap is directly related to the possibility of the band inversion which is the key topological property\cite{bernevig2006d}. For example, sometimes LDA and GGA predict a negative band gap, whereas the band gap is positive in reality\cite{Perry2001}. This can cause serious problem to predict topological insulators. So it is necessary to improve the calculations of the energy gap. The most effective method to calculate the band bap is GW approximation\cite{Louie-GW1986}. Simply saying, GW approximation considers the Hartree-Fock self-energy interaction with the screening effect. Though the GW method has been used to study topological insulators\cite{Yazyev2012}, this method is very expensive. Besides the GW method, the modified Becke-Johnson exchange potential together with LDA(MBJLDA), proposed by Tran and Blaha in 2009\cite{Tran2009}, costs as much as LDA and GGA, but it allows the band gap with the similar accuracy to GW's. MBJLDA potential can also recover LDA for the electronic system with a constant charge density, and mimic the behavior of orbital-dependent potentials as well.

LDA+U\cite{Anisimov1991}, LDA+DMFT\cite{Georges1996} and LDA+Gutzwiller\cite{Deng2009} are employed to study to strongly correlated electronic systems(\textbf{d} and \textbf{f} electrons), because LDA often fails for these systems. In strongly correlated electronic systems, the electrons are strongly localized, and have more features of atomic orbitals. This case requires proper treatment of atomic configurations and orbital dependence. Both LDA and GGA don't include the orbital-dependent feature, this is why they fail to describe strongly correlated electronic systems. Based on this understanding, all of LDA+U, LDA+DMFT and LDA+Gutzwiller include the orbital-dependent feature in different ways. For example, the on-site interaction is treated in a static Hartree mean-field manner in LDA+U method which is simplest and cheapest method. It is often used for strongly correlated systems, but it does not work well with intermediately correlated metallic systems. The self-energy of the LDA+DMFT method is obtained in a self-consistent way. Up to now LDA+DMFT is most accurate and reliable method, but its computational costs are high. LDA+Gutzwiller based on Gutzwiller variational approach is recently developed. This method works well for intermediately correlated electronic systems, and it is cheaper than LDA+DMFT. Though it is still an open question how well these methods work on strongly correlated systems, it is true that these methods could reproduce some results of experiments, and that they can help to understand some novel results in strong correlated electronic systems.

\subsection{Spin-orbit coupling}
Generally the SOC describes the interaction of a particle's spin with its orbital motion. For example, in one atom, the interaction between one electron's spin and the magnetic field produced by its orbit around the nucleus can cause shifts in the electron's atomic energy levels, which is the typical SOC effect. In the non-relativistic limit, SOC Hamiltonian from relativistic Dirac equation is written as\cite{winkler2003},
\begin{equation}\label{soc}
    H_{soc}=-\frac{\hbar}{4m_0^2c^2}\sigma\cdot\textbf{p}\times(\nabla V_0)
\end{equation}
where $\hbar$ is Planck's constant, $m_0$ is the mass of a free electron, $c$ is the velocity of light and $\sigma$ represents the Pauli spin matrices.
$H_{soc}$ couples the potential $V_0$ and the momentum operator \textbf{p} together.

In the case of the single atomic system $V_0$ is spherically symmetric, $H_{soc}$ can be simplified,
\begin{equation}\label{atomic-soc}
   H_{soc}=\lambda\textbf{L}\cdot\sigma
\end{equation}
where $\lambda$ is the strength of SOC interaction. \textbf{L} represents the angular moment. But in solid systems, $V_0$ is the periodic potential which can be very complex, so it is difficult to exactly calculate SOC effect. For this case, SOC effect is usually calculated with a second-variational procedure. SOC interaction is the key to the band topology, so all first-principles calculations to study topological insulators should be carried out with SOC.

\subsection{The criterion of topological insulators}
There are four $Z_2$ invariants($\nu_0$;$\nu_1$$\nu_2$$\nu_3$) for three dimensional topological insulators, first proposed by Fu, Kane and Mele\cite{fu2007b}. When $\nu_0=1$, states are strong topological insulators which have topologically protected gapless surface states consisting of odd Dirac cones. These surface state are robust against time-reversal-invariant(TRI) weak disorders. If $\nu_0=0$ and at least one of $\nu_{1,2,3}$ is non zero, the corresponded states are weak topological insulators which have surface states with even Dirac cones on special surfaces. We can simply consider weak topological insulators to be stacked by layered two-dimensional QSH states. In the presence of disorders, the surface states of weak topological insulators can be destroyed. When all $\nu_{0,1,2,3}$ are zero, states are conventional insulators.

\subsubsection{With the inversion symmetry}
 The calculation of $Z_2$ invariants is very simple for the compounds with the inversion symmetry. The formula of $Z_2$ can be just expressed with the parity values at the eight time-reversal-invariant moments(TRIMs)\cite{fu2007a},
\begin{equation}\label{v0-inversionsymmetry}
    (-1)^{\nu_0}=\prod_{i=1}^8\delta_i
\end{equation}
and
\begin{equation}\label{v123-inversionsymmtry}
    (-1)^{\nu_k}=\prod_{n_k=1;n_{j\neq k}=0,1}\delta_{i=(n1,n2,n3)}
\end{equation}
where
  \begin{equation}\label{parity}
    \delta_i=\prod_{m=1}^N\xi_{2m}(K_i)
\end{equation}
$N$ is half of the number of occupied bands, and $\xi_{2m}(K_i)$ is the parity eigenvalue of the $2m$th occupied energy band at TRIM $K_{i=(n_1n_2n_3)}=\frac{1}{2}(n_1\mathbf{b_1}+n_2\mathbf{b_2}+n_3\mathbf{b_3})$ where $\mathbf{b_{1,2,3}}$ represent primitive reciprocal-lattice vectors.

\subsubsection{Without the inversion symmetry}
For the compounds without the inversion symmetry, $Z_2$ invariants also can be calculated by the numerical method proposed by Fukui \emph{et al} based on the tight-binding method\cite{Fukui2007}. Firstly, $Z_2$ formula of QSH state can be expressed with the Berry connection and the Berry curvature, shown by Fu and Kane,
\begin{equation}\label{QSH-Z2}
    Z_2=\frac{1}{2\pi}[\oint_{\partial \mathcal{B}  ^-}A(k)-\int_{ \mathcal{B} ^-}F(k)]  \text{ mod 2}
\end{equation}
with
\begin{equation}\label{berry}
A(k)=i\Sigma_n\langle u_n(k)|\nabla_k u_n(k)\rangle \text{~and~} F(k)=\nabla_k\times A(k)
\end{equation}
where $\mathcal{B}^-$ and $\partial \mathcal{B}^-$ indicate half of two-dimensional(2D) tori and its boundary, respectively. In order to do numerical calculations, equation (\ref{QSH-Z2}) can directly be rewritten to its lattice version. Secondly, for 3D case, we can define six 2D tori as Z$_0$($k_x$,$k_y$,0), Z$_1$($k_x$,$k_y$,$\pi$), Y$_0$($k_x$,0,$k_z$), Y$_1$($k_x$,$\pi$,$k_z$), X$_0$(0,$k_y$,$k_z$) and X$_1$($\pi$,$k_y$,$k_z$). We can calculate the $Z_2$ based on equation (\ref{QSH-Z2}) for each of these six tori, as $z_0$, $z_1$, $y_0$, $y_1$, $x_0$ and $x_1$. The four $Z_2$ invariants of topological insulators are obtained by $\nu_0=x_0x_{\pi}$, $\nu_1=x_{\pi}$, $\nu_2=y_{\pi}$ and $\nu_3=z_{\pi}$. Xiao \emph{et al.} first successfully used these formulas to evaluate the $Z_2$ invariants of half-Heusler compounds by first-principles calculations\cite{Xiao2010}.

\subsubsection{Adiabatic argument}
Sometimes it is not necessary to directly calculate $Z_2$ for the compounds without inversion symmetry. One can start from an according compound with the inversion symmetry, and then adiabatically change this compound to that without inversion symmetry. If the energy gap does not close in an adiabatic process, the topological property will not change. For example, the space group of $\alpha$-Sn is Fd$\overline{3}$m(No. 227) and the inversion symmetry is held in this structure. We can easily know $\alpha$-Sn is topologically nontrivial from the parity calculations\cite{fu2007a}. Based on the adiabatic argument, one can conclude HgTe is topologically nontrivial.

\subsubsection{Surface states}
Gapless surface states of topological insulators must include the odd number of Dirac cones on one surface, and these surface states are robust against TRI weak disorders. So the calculation of surface states is another useful method to judge the band topology. The simplest way to calculate surface states is based on the free-standing structure. It is true that this is very powerful method to calculate surface states, but only for the compounds with the inversion symmetry and layered structure, such as, Bi, Sb, Bi$_2$Se$_3$ and so on. For example, if the compounds do not have the inversion symmetry, the polarization field might cause serious artificial effect, especially for the compounds with a small band gap. In addition, if the compounds are not layered structure, the dangling bonds on the surface might cause a number of complex topologically trivial chemical surface states which can mix with topologically non-trivial ones. The topological surface states originate from the topological property of the bulk electronic structure. Though the details of these surface states can be modified by the special dangling bonds and the reconstruction of the electronic structure on the surface, we address that the topological feature don't change, such as, the odd number of Dirac cones. The calculation of the free-standing model also costs a lot, because the vacuum layer should be thick enough in order to avoid the hybridization between the up and down surfaces.

 Besides the free-standing model, maximally localized Wannier function(MLWF) methods\cite{marzari1997,souza2001} can be used to calculate the surface states\cite{zhang2009,Zhang-BiSb2009}. Essentially the MLWF method is a tight-binding method, but the difference from the conventional tight-binding method is that MLWF method can exactly reproduce the band structure of first-principles calculations. But it is not easy to obtain MLWFs, because the transformation from Bloch functions to Wannier functions is not unique due to the phase ambiguity of the Bloch functions used in first-principles calculations. Marzari and Vanderbilt reported an effective method to obtain MLWF by minimizing the spread function $\sum_n(\langle \mathbf{r}^2\rangle-\langle \mathbf{r}\rangle ^2)$\cite{marzari1997}. In order to calculate surface states, first we carry out the first-principles calculations for 3D bulk structure and then transform Bloch functions to MLWFs. At the same time the hopping parameters $H_{mn}(\mathbf{R})=\langle n0|\hat{H}|m \mathbf{R}\rangle$ between Wannier functions are obtained. At the next step, we use these hopping parameters to construct the hopping parameters of the corresponding semi-infinite structure, and then iterative method can be used to solve the surface Green's function,
\begin{equation}\label{sgreen}
    G_{nn}^{\alpha,\alpha}(\mathbf{k_{||}},\epsilon+\mathbf{i}\eta)
\end{equation}
Where $n$ denotes the unit cell along the surface norm, and $\alpha$ is the Wannier orbital in the unit cell. The MLWFs method can predict surface states well for layered compounds. For example, the calculated surface states of Bi$_2$Se$_3$ with MLWFs method agree well with the ones of Angle Resolved Photoelectron Spectroscope(ARPES)\cite{zhang2009,Xia2009}. Usually we don't expect to predict the exact dispersion of surface states, because this method does not include all complex situations on the surface. On the other hand, the surface states obtained from the MLWF method originate from the topological property of the bulk electronic structure, so this is an ideal method to judge whether one compound is topologically non trivial or not.

\begin{figure}
  \center
  \includegraphics[width=.45\textwidth,angle=-90,clip]{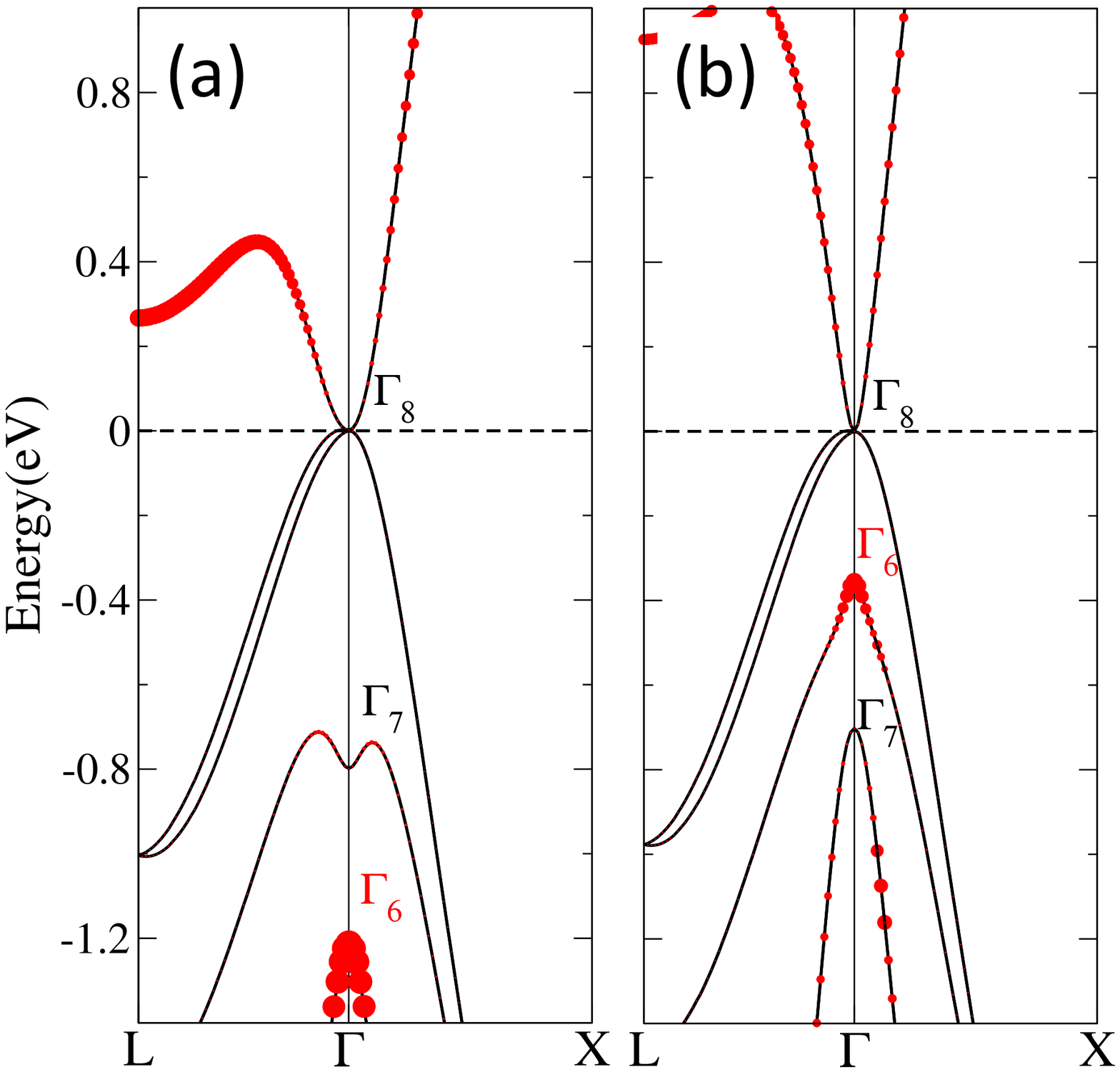}\\
  \caption{(a) and (b) are the band structure of HgTe by LDA and MBJLDA methods, respectively. $\Gamma_{6,7,8}$ represent the symmetry of energy levels at $\Gamma$ point. The solid red circles indicate the projection of the \textbf{s} orbital of Hg. The LDA band structure shows the $\Gamma_8-\Gamma_7-\Gamma_6$ band sequence which is not correct, but MBJLDA can calculate the correct band sequence as $\Gamma_8-\Gamma_6-\Gamma_7$. }\label{HgTeband}
\end{figure}

\begin{figure*}
  \center
  \includegraphics[width=.46\textwidth,angle=-90,clip]{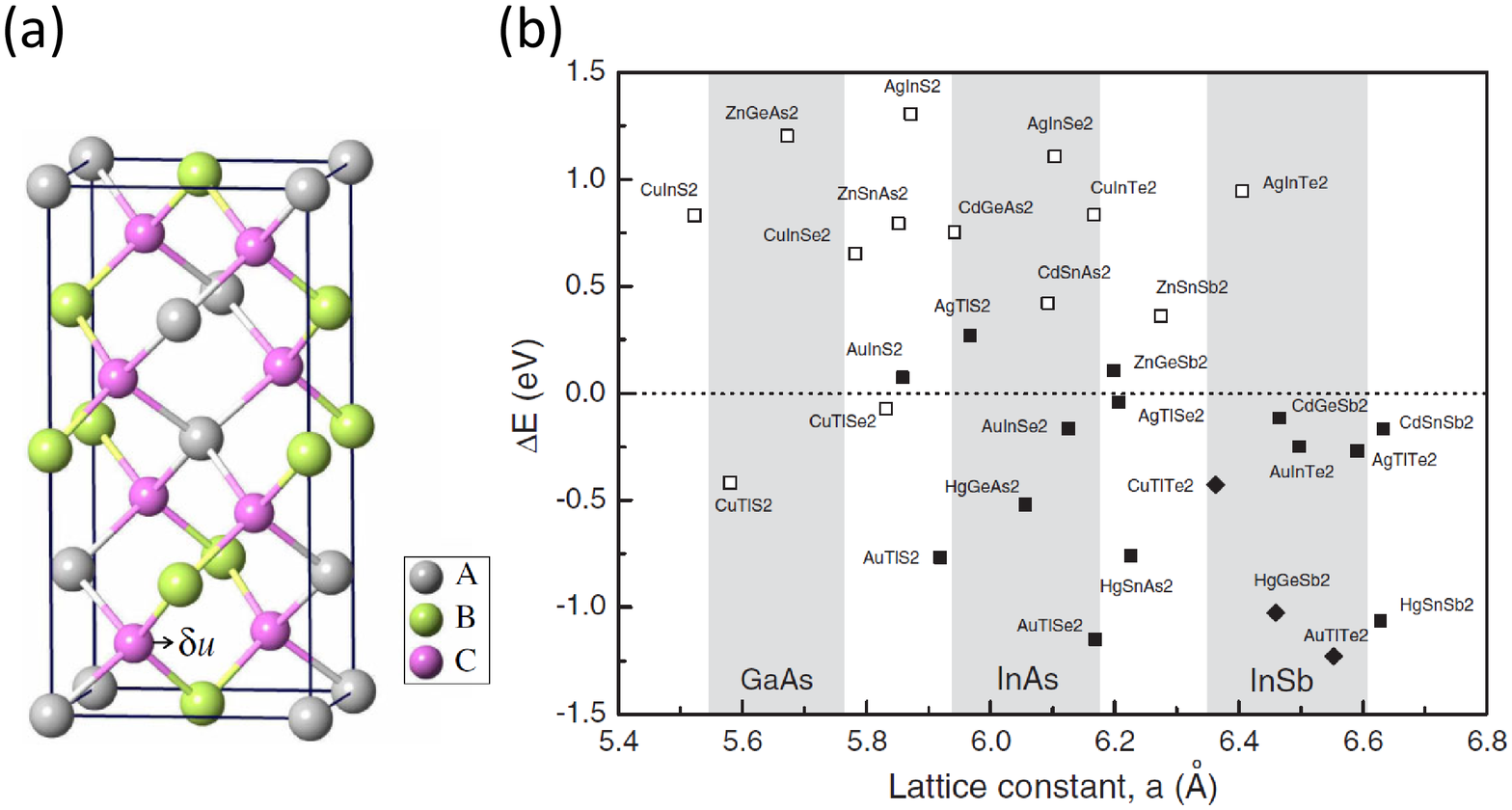}\\
  \caption{(a) shows the crystal structure of chalcopyrite compounds(ABC$_2$).(b) shows the energy gap $\Delta E$ for various chalcopyrite compounds as a function of the lattice constant. Open symbols mean the lattice constant has been reported. The lattice constants of the rest are obtained by first-principles total energy minimization. Squares represent topological insulators, and diamonds represent topological metals. From Ref\cite{Feng2011} }\label{Feng}
\end{figure*}

\begin{figure*}
  \center
  \includegraphics[width=.45\textwidth,angle=-90,clip]{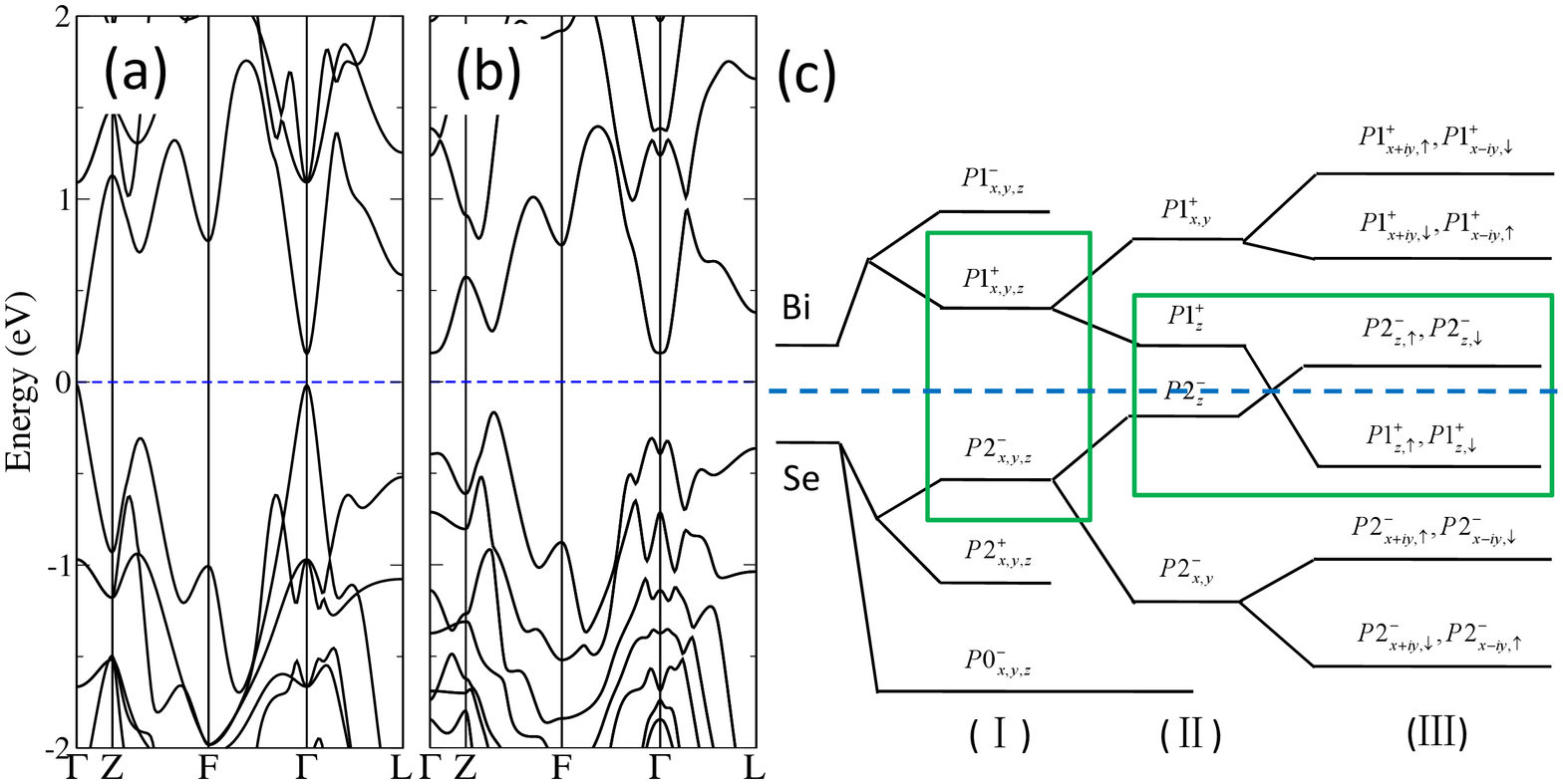}\\
  \caption{The band structure of Bi$_2$Se$_3$ without (a) and with (b) SOC. The blue dashed represents Fermi level. (c), the evolution of the band sequence at $\Gamma$ point starting from atomic levels. The three stage (\textrm{I}), (\textrm{II}) and (\textrm{III}) represent turning on chemical bonding, crystal field and SOC effects step by step.  From Ref\cite{zhang2009} }\label{Bi2Se3band}
\end{figure*}

\begin{figure*}
  \center
  \includegraphics[width=.43\textwidth,angle=-90,clip]{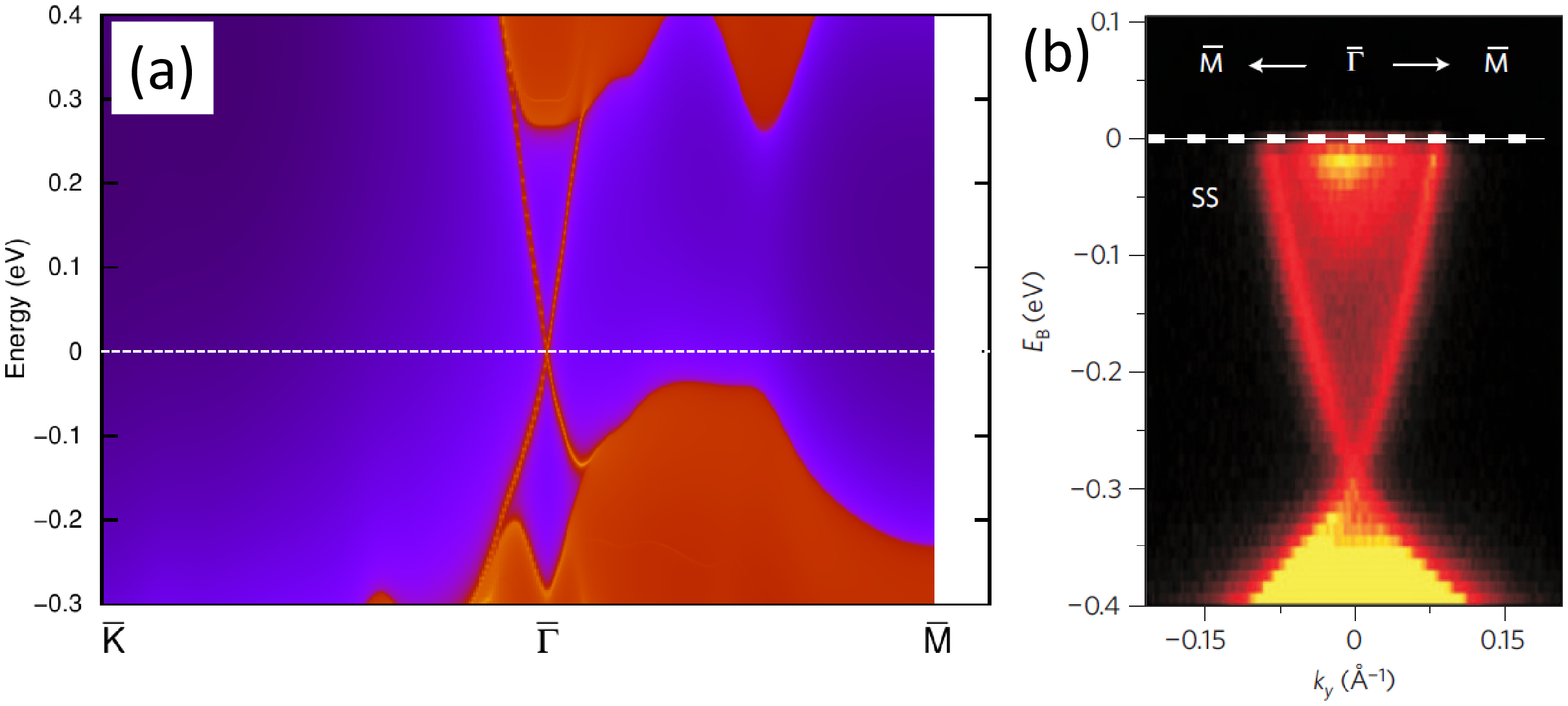}\\
  \caption{(a), the calculated surface states of Bi$_2$Se$_3$ with MLWFs tight-binding method for a semi-infinite structure with the surface normal(111). The more red means more states, and the blue region indicates band gap without any states. The clear surface states with the linear dispersion at $\overline{\Gamma}$ can be seen in the band gap.(b), the ARPES result for Bi$_2$Se$_3$ along the $\overline{\Gamma}-M$ dirction.  From Ref\cite{zhang2009,Xia2009} }\label{Bi2Se3ss}
\end{figure*}

\section{3-dimensional topological insulators}
After the initial discovery of the 2D topological insulator HgTe\cite{bernevig2006d,koenig2007}, a number of 3D topological insulators are found with the great effort of theorists and experimentalists\cite{Qi-physics-today2010,Hasan2010,Qi2011}. In the following, we classify the topological insulators by the type of the band inversion, because the band inversion has a clear and general physical picture for most topological insulators. Up to now, there are three basic types of band inversions (\textbf{s}-\textbf{p}, \textbf{p}-\textbf{p}, \textbf{p}-\textbf{f}) in topological insulators discovered so far. In the following discussions, we will take some representative compounds as examples for each type of topological insulators.

\subsection{s-p type}
The most important \textbf{s}-\textbf{p} topological insulator is HgTe\cite{bernevig2006d,koenig2007} which has the zinc-blende structure with the space group F$\overline{4}3m$(No.216). Before HgTe was found to be a topologically non-trivial compound, it had been widely studies experimentally and theoretically\cite{Capper1987,Lu-HgTe1989,Delin2002}. Unlike other zinc-blende compounds, HgTe is a semiconductor with the symmetry-protected zero energy band gap . The Hg has occupied shallow \textbf{5d} levels which tends to be delocalized, so Hg has a large effective positive charge in its core. The Hg's \textbf{s} level, which forms $\Gamma_6$ state in cubic symmetry, is pulled down below the Te's \textbf{p} levels which split into $\Gamma_8$ and $\Gamma_7$, by this effective positive charge of Hg's core. Finally the energy level sequence at $\Gamma$ point shows the $\Gamma_8-\Gamma_6-\Gamma_7$ order, which we call the \textbf{s}-\textbf{p}-type band inversion. If we define the energy gap $\Delta E$,
\begin{equation}\label{negativegap}
    \Delta E=E_{\Gamma_6}-E_{\Gamma_8}
\end{equation}
where the $E_{\Gamma_6}$ and $E_{\Gamma_8}$ are the energy levels for $\Gamma_6$ and $\Gamma_8$ at the $\Gamma$ point. HgTe has a negative $\Delta E$ because the \textbf{s}-\textbf{p}-type band inversion, so it is well known as a negative gap semiconductor.

The normal LDA and GGA can predict the band inversion between $\Gamma_6$ and $\Gamma_8$, but the exact band sequence of $\Gamma_8-\Gamma_6-\Gamma_7$ cannot be obtained\cite{Delin2002}. The LDA band structure with SOC shows the $\Gamma_8-\Gamma_7-\Gamma_6$ sequence, shown in fig\ref{HgTeband}(a). As we addressed in the above, MBJLDA method can correct the error of LDA band structure. The band structure with the MBJLDA method is shown in fig\ref{HgTeband}(b), which perfectly shows the correct $\Gamma_8-\Gamma_6-\Gamma_7$ sequence.

Bernevig, Hughes and Zhang first identified the band inversion in HgTe to be the key ingredient of its topologically non-trivial behavior\cite{bernevig2006d}. Its topological invariant can also be obtained by an adiabatic argument\cite{fu2007a}. As we know, if we replace Hg and Te by the same atom in the zinc-blende structure, the crystal structure will change to the diamond structure with the inversion symmetry. Luckily, in nature grey tin has the diamond structure with space group F$d\overline{3}m$, and it also is a semiconductor with a negative energy gap $\Delta E$ due to the \textbf{s} level below the \textbf{p} level.  Because grey tin holds the inversion symmetry, its parity values at all TRIMs can be easily calculated. It is worth to note that though grey tin is a zero band gap semiconductor, we still can define the topological property for all of its occupied bands. Based on the formulas proposed by Fu and Kane, its $Z_2$ invariants are calculated to be (1;000) which indicate topologically non trivial. Here the key is that the \textbf{s} and \textbf{p} at $\Gamma$ point have opposite parity values. The occupied \textbf{s} state forms $\Gamma_7^-$, whereas \textbf{p} states form $\Gamma_7^+$ and $\Gamma_8^+$. Taking grey tin as the starting point, we can make a thought experiment to adiabatically change grey tin to HgTe. In this process, the negative gap($\Delta E$) is never closed, which means the grey tin and HgTe have the same topological property. So HgTe prove to be topologically non trivial with $Z_2$ invariant(1;000). Besides this adiabatic argument, HgTe's $Z_2$ invariants can also be directly calculated by the numerical method addressed above.

Similar to HgTe, there are a big family compounds known as half-Heusler materials(XYZ)\cite{Felser2007} which include more than 250 semiconductors and semimetals. Half-Heusler compounds have the face-centered cubic(fcc) structure sharing the same space group with HgTe. Y and Z form zinc-blende structure which is stuffed by X. Usually X and Y are transition metal or rare earth elements, and Z is a heavy element. The band structure of half-Heusler compounds at $\Gamma$ point near the Fermi level is almost the same with that of HgTe case. \textbf{s} state forms $\Gamma_6$, and \textbf{p} states split into $\Gamma_7$ and $\Gamma_8$. Some of half-Heusler compounds have the band sequence $\Gamma_6-\Gamma_8-\Gamma_7$ and some others have the inverted band sequence $\Gamma_8-\Gamma_6-\Gamma_7$. The interesting thing is that half-Heusler family were independently reported almost at the same time by three theory groups\cite{Chadov2010,Lin2010,Xiao2010}. Besides the topological property, half-Heusler compounds are a class of multifunctional materials\cite{Canfield1991,Goll2008}, such as, superconductivity and magnetism, due to transition metals and rare earth elementals. So half-Heusler compounds might be the best platform to study the Majorana fermion in topological superconductors\cite{Qi-TSC2009}, dynamical axion field in topological anti-ferromagnetic phase\cite{Li2010}, and quantum anomalous Hall effect(QAH) in topological ferromagnetic phase\cite{Yu2010}. Recently, some experiments, such as, ARPES and transports, already have been reported for half-Heusler compounds\cite{Gofryk-Heusler-transport2011,Liu-Heusler-ss2011,Shekhar-Heusler-transport2012}.

Generally due to the cubic symmetry, many topologically non trivial compounds(HgTe and half-Heusler compounds) are zero gap semiconductors with Fermi level through $\Gamma_8$ level at $\Gamma$ point, and a uniaxial strain is usually needed to break the cubic symmetry in order to open a finite energy gap\cite{dai2008}. Feng \emph{et al.} reported that chalcopyrite structure can naturally break the cubic symmetry\cite{Feng2011}. The chalcopyrite structure(ABC$_2$) is the body-centered tetragonal structure  with the space group I$\overline{4}2d$(No. 122), which could be regarded as a superlattice of two cubic zinc-blende unit cell AC and BC, seen in fig\ref{Feng}(a). In essence, the unit cell of chalcopyrite are the double unit cell of HgTe with naturally breaking the cubic symmetry, and we expect that these two class compounds might share the same topological property. Feng \emph{et al.} found that it is true that some materials with chalcopyrite structure are topological insulators, shown in fig\ref{Feng}(b).

Besides the compounds talked about above, there are a lot of other \textbf{s}-\textbf{p}-type topological insulators, such as, $\beta-$Ag$_2$Te\cite{Zhangwei2011}, KHgSb family\cite{Zhang2011}, Na$_3$Bi\cite{Wang2012}, CsPbCl$_3$ family\cite{Yang2012} and so on.

\begin{figure}
  \center
  \includegraphics[width=.5\textwidth,angle=0,clip]{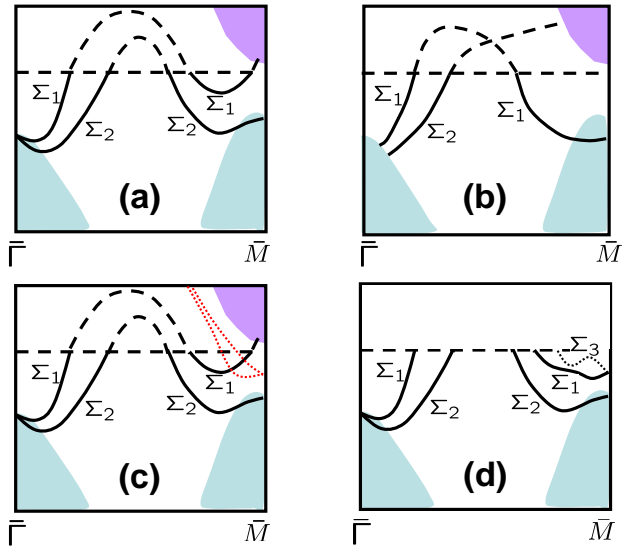}\\
  \caption{ Schematic for the comparison of the surface states of (a) first-principles calculations\cite{Zhang-BiSb2009}, (b) tight-binding calculations\cite{Teo2008} and (c) ARPES experiment\cite{hsieh2008}. From Ref\cite{Zhang-BiSb2009} }\label{BiSbss}
\end{figure}

\subsection{p-p type}
Due to the simple surface states consisting of a single Dirac cone, Bi$_2$Se$_3$, Bi$_2$Te$_3$ and Sb$_2$Te$_3$ compounds\cite{zhang2009,Xia2009,Moore2009,Chen2009,Zhangyi2010,Peng2010} quickly became topological insulators extensively studied worldwide. Especially Bi$_2$Se$_3$ has a big energy gap of 0.3eV which is much larger than the energy scale of room temperature. These compounds share the layered structure with a five-atom layer, called the quintuple layer(QL), as the unit cell with the space group R$\overline{3}m$(No. 166). Two equivalent Se atoms, two equivalent Bi atoms and a third Se atom are in each QL. The coupling is the chemical bonding between neighbor atomic layers whin one QL, but the van der Waals type, which is much weaker, between two QLs. It is worth to note that the inversion symmetry is held in the crystal structure.

In the following, we briefly introduce the basic electronic structure of this family compounds by taking Bi$_2$Se$_3$ as an example. First of all, the band structure without SOC shows Bi$_2$Se$_3$ to be a narrow band gap insulator. Both the bottom of conduction band and the top of valence band are at $\Gamma$ point, seen in fig\ref{Bi2Se3band}(a). After SOC is turned on, the bottom of conduction band is pulled down below the top of valence band, and an interaction gap opens at the crossing of valence band conduction bands, seen in fig\ref{Bi2Se3band}(b). Based on the parity calculations, $Z_2$ invariants of Bi$_2$Se$_3$ are calculated to be (1;000) which mean topologically non trivial. The key to be topological insulators is that the bottom of conduction band and the top of valence band have opposite parity values. The schematic of the band sequence at $\Gamma$ point clearly tells the band evolution starting from atomic levels with three stages, shown in fig\ref{Bi2Se3band}(c). Because the \textbf{s} levels are much lower than \textbf{p} levels, we just start from the atomic \textbf{p} levels of Bi(6\textbf{s}$^2$6\textbf{p}$^3$) and Se(4\textbf{s}$^2$4\textbf{p}$^4$). At the stage(I), the bonding and anti-bounding effect between Bi and Se atoms are considered. All the atomic orbitals are recombined into $P0_{x,y,z}^-$, $P1_{x,y,z}^\pm$ and $P2_{x,y,z}^\pm$ where '$0$' represents the third Se, and '$1$', '$2$' represent Bi and the other two Se, respectively. '$\pm$' represents the parity values. At the stage (II), after the crystal field is turned on,  $\mathbf{p}_{xyz}$ levels will split into $\mathbf{p}_{xy}$ and $\mathbf{p}_z$. The levels of $P1_{z}^+$ and $P2_{z}^-$ are nearest to Fermi level. At the stage (III), SOC effect is further introduced. $P1_{z}^+$ becomes two degeneracy levels($P1_{z,\uparrow\downarrow}^+$), and $P2_{z}^-$ becomes two degeneracy levels($P2_{z,\uparrow\downarrow}^-$) due to the time-reversal symmetry. SOC effect tries to pull $P1_{z,\uparrow\downarrow}^+$ down and push $P2_{z,\uparrow\downarrow}^-$ up. Finally if SOC is strong enough, the \textbf{p}-\textbf{p}-type band inversion will happen between $P1_{z,\uparrow\downarrow}^+$ and $P2_{z,\uparrow\downarrow}^-$.

Due to the layered structure with inversion symmetry, both the free-standing model and the tight-binding model based on MLWFs can be used to calculate surface states. Fig\ref{Bi2Se3ss}(a) shows the clear surface states of Bi$_2$Se$_3$ with a single Dirac cone at $\overline{\Gamma}$ calculated by the MLWFs tight-binding model. Almost at the same time of Zhang $et~al.$'s theory prediction\cite{zhang2009}, Hasan group reported the the topologically non-trivial surface states of Bi$_2$Se$_3$ by the ARPES experiment\cite{Xia2009}, shown in fig\ref{Bi2Se3ss}(b). Comparing the theory and experiment results, we have to agree that first-principles calculations can successfully predict topological insulators, including the details of surface states. Recently a lot of experimental studies of topological insulators are focusing on these compounds, because these compounds are easily to be grown by all kinds of experiments.

The topological insulator Bi$_{1-x}$Sb$_x$($0.07<x<0.22$) alloy also belongs to the \textbf{p}-\textbf{p} type\cite{fu2007a}. Bulk Bi and Sb share a rhombohedral R$\overline{3}m$ structure which holds the inversion symmetry, and they both are semimetals with some tiny Fermi pockets around the TRIM $L$ and $T$ points, but there is a direct gap at every \textbf{k} point through the whole Brillouin zone(BZ). So we can define an imaginary Fermi surface in the direct gap. Based on the parity calculations, we confirm that Bi is topologically trivial with $Z_2$ (0;000), and that Sb is topologically non trivial with $Z_2$ (1;111). The key difference of Bi and Sb is the band sequence of the conduction and valence bands at three $L$ points. For example, the conduction band of Bi is $L_s$, and the valence band is $L_a$ where '$a/s$' indicate the $-/+$ parity. Differently, these two bands switched with each other in Sb. After carefully comparing the band structure between Bi and Sb, Fu and Kane predicted that the insulate phase of Bi$_{1-x}$Sb$_x$ ($0.07<x<0.22$) alloy must be a topological insulator. Subsequently, Hasan group observed the topologically non-trivial property of Bi$_{1-x}$Sb$_x$ by the ARPES experiment\cite{hsieh2008}. But the details of surface states don't agree with ones of tight-binding\cite{Teo2008} and first-principles calculations\cite{Zhang-BiSb2009}. The schematic of the difference among these results are shown in fig\ref{BiSbss}.  We can see that the ARPES result indicates three surface states $\sum_{1,2,3}$, but two surface state $\sum_{1,2}$ are only found by tight-binding and first-principles calculations. Zhang $et al.$ argued that the extra surface state $\sum_3$ might come from the imperfect surface, but this still is an open question up to now.

Following Bi$_2$Se$_3$ family, a number of other \textbf{p}-\textbf{p} Bi-based topological insulators are predicted by theories and observed by experiments, such as, TlBiSe$_2$ family\cite{yan2010}, SnBi$_2$Te$_4$ and SnBi$_4$Te$_7$ family\cite{Eremmev2012}, and so on.

\begin{figure}
  \center
  \includegraphics[width=.45\textwidth,angle=-90,clip]{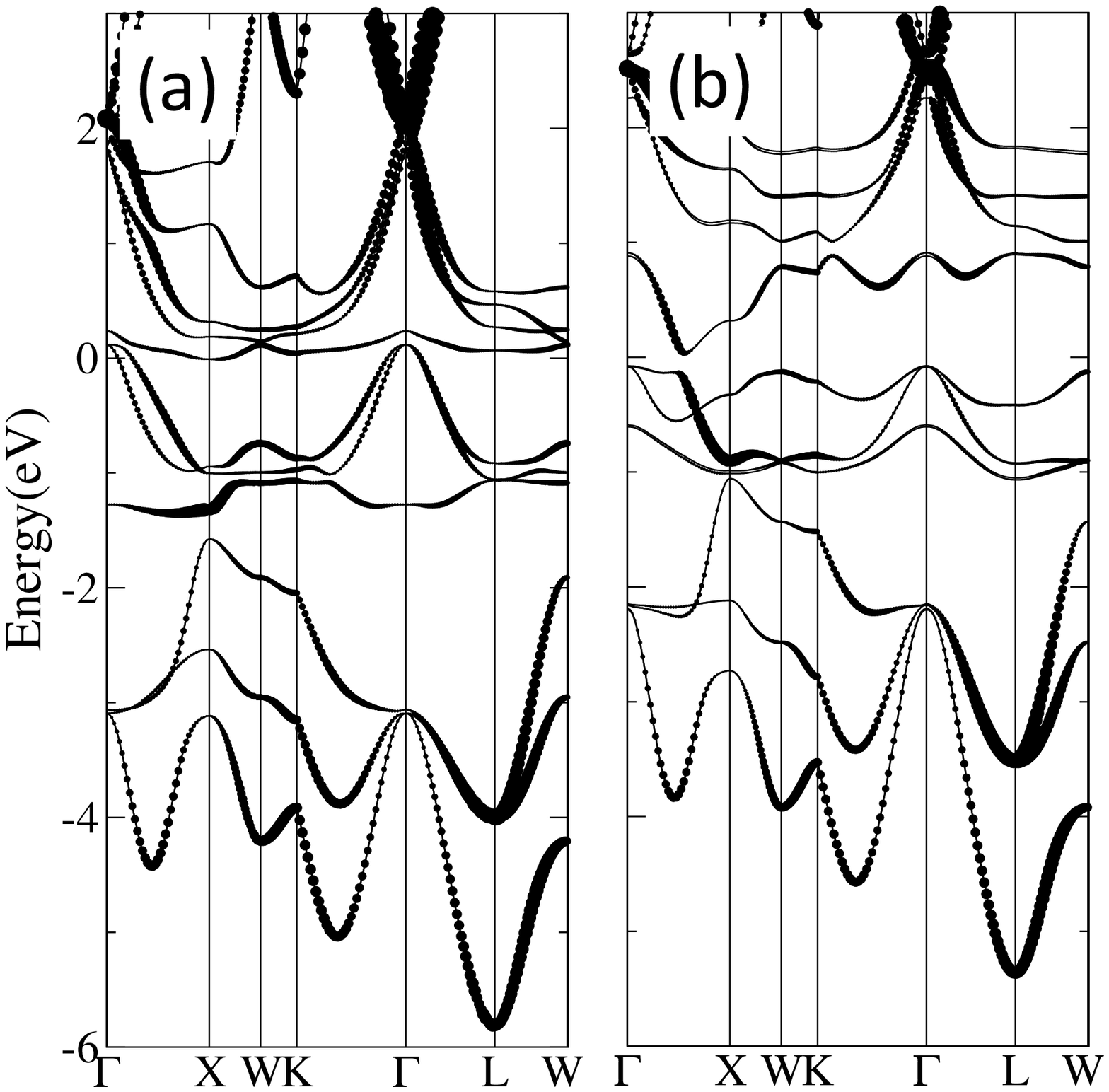}\\
  \caption{The band structure for AmN compound with U=0eV (a) and U=2.5eV (b) from the LDA+U method. The thickness of the band is corresponded to the projected weight of the \textbf{d} character of Am. In $\Gamma-X$ direction, one band with\textbf{d} character clearly comes down to cross with the valence bands which mainly have the \textbf{f} character. (a) represents semi metal without a band gap, but (b) represents a finite band gap. From Ref\cite{Zhang2012} }\label{AmNband}
\end{figure}

\subsection{d-f type}
There is no clear evidence for the limit of the energy gap size for topological insulators. How could we find new topological insulators with bigger energy gap($>$0.3eV)? One possible way to enhance the SOC energy gap is to consider the cooperation of the SOC interaction and other effects, such as, the electron-electron correlation. In this idea, topological Kondo insulators were proposed, and SmB$_6$ as an example was predicted to be a topological Kondo insulator\cite{Dzero-KondoTI2010}. Though due to 4\textbf{f} orbitals SmB$_6$is a strong correlated system, it only has a tiny energy gap. Recently Zhang \emph{et al.} predicted AmN and PuTe family compounds are \textbf{d} and \textbf{f} topological insulators with strong interaction\cite{Zhang2012}. All AmN and PuTe family compounds have rock-salt crystal structure with the space group F$m\overline{3}m$(No. 225), and the inversion symmetry is also held in this structure. All these compounds have been well studied by theories and experiments, known as mixed valence materials. Here we take AmN as an example to understand the band structure. The configuration of actinide Am is 5\textbf{f}$^7$7\textbf{s}$^2$6\textbf{d}$^0$. The SOC interaction is strong than Hund's rule, so the \textbf{f} orbitals split into high energy $J=7/2$ and low energy $J=5/2$ states. Approximately, in AmN, Am forms Am$^3+$ with the configuration 5\textbf{f}$^6$7\textbf{s}$^0$6\textbf{d}$^0$, the states of $J=5/2$ should be fully occupied, and $J=7/2$ states are unoccupied. But due to the delocalization of 5\textbf{f} in Am, 5\textbf{f} states partly hybridize with 6\textbf{d} states with neighbor Am atoms.

In the fcc crystal field, \textbf{d} orbitals first split into $t_{2g}$ and $e_g$ states, and $t_{2g}$ level goes down to cross 5\textbf{f} below Fermi level along $\Gamma-X$ direction, shown in fig\ref{AmNband}. The band inversion happens at three $X$ points. If only LDA calculations are used, the full energy gap cannot open through the whole BZ. After the electron correlation is introduced with LDA+U method, a band gap can open up with the proper correlation parameter U. We have to address that the electron correlation U is found to enhance the SOC in these compounds. Because there are three TRIM $X$ point in BZ, $Z_2$ invariants of AmN must be topologically non trivial. Further more, our conclusion suggests all the mix-valence compounds with rock-salt structure must be topologically non-trivial. Especially, there transport experiments showed PuTe\cite{Ichas2001} has a big energy gap around 0.2eV, and this gap can be enhanced to 0.4eV with the pressure. Many of these \textbf{f} compounds host all kinds of magnetic phases, so they might open the opportunity to study QAH effect and dynamic Axion field.

\section{Summary and outlook}
In this review, we first introduced widely-used techniques within first-principles calculations including LDA and GGA, GW and MBJLDA, LDA+U, LDA+DMFT and LDA+Gutzwiller methods, because they play a crucial role on the field of topological insulators. Then the basic concepts of topological insulators and some useful methods to confirm the topological property are summarized. We classify topological insulators found to-date into three types as  \textbf{s}-\textbf{p}, \textbf{p}-\textbf{p} and \textbf{d}-\textbf{f} based on the clear band inversion picture. For each type of topological insulators, we take several typical compounds as examples with talking about the electronic structure and the topological property.

Though many topological insulators have been discovered, it is still important to find more with desired properties. First of all, a big band gap is important for the application of surface states of topological insulators. Up to now the biggest band gap is around 0.3eV in Bi$_2$Se$_3$ compound. Secondly, the transport experiments to detect surface states are still very challenging\cite{Veldhorst2012,Kim2012,Hong2012}. One reason is that the quality of samples is not good enough with a low mobility. Another reason is that Dirac cone always coexists with some bulk carriers. In order to overcome this barrier, on the one hand, experimentalists are trying to improve the quality of samples. On the other hand, it is important to find other new topological insulators with functional properties. In addition, it is interesting to study the cooperation of the topological property with other phases, such as, superconductivity, magnetism and so on. We hope that this review can provide some guidance in the search.

\begin{acknowledgement}
This work is supported by the Defense Advanced Research Projects Agency Microsystems
Technology Office, MesoDynamic Architecture Program (MESO) through the contract number
N66001-11-1-4105 and by the Army Research Office (No.W911NF-09-1-0508).
\end{acknowledgement}

%

%

\providecommand{\WileyBibTextsc}{}
\let\textsc\WileyBibTextsc
\providecommand{\othercit}{}
\providecommand{\jr}[1]{#1}
\providecommand{\etal}{~et~al.}

%
%
%
%
%
%
%

\end{document}